\documentclass[journal]{IEEEtran}

\usepackage{ifpdf}
\ifCLASSINFOpdf
  \usepackage[pdftex]{graphicx}
\else
   \usepackage[dvips]{graphicx}
\fi
\usepackage{cite}
\usepackage{amsmath}
\usepackage{algorithmic}
\usepackage{array}
\usepackage{enumerate}
\usepackage{color}
\usepackage{multicol}
\usepackage{multirow}
\usepackage{graphicx}
\usepackage{epstopdf}
\usepackage{caption}
\usepackage{subfigure}
\usepackage{amsfonts}

\begin{document}
\captionsetup[figure]{labelformat={default},labelsep=period,name={Fig.}}
\captionsetup[table]{font=small, justification=centering, labelsep=newline, textfont=sc}
\renewcommand{\thetable}{\Roman{table}}

%
\title{PulseGAN: Learning to generate realistic pulse waveforms in  remote photoplethysmography}
\author{Rencheng~Song,~\IEEEmembership{Member,~IEEE,}
       Huan~Chen,
        Juan~Cheng,~\IEEEmembership{Member,~IEEE,  }
              Chang~Li,~\IEEEmembership{Member,~IEEE,}
                Yu~Liu,~\IEEEmembership{Member,~IEEE,}
        and~Xun~Chen,~\IEEEmembership{Senior Member,~IEEE}

\thanks{R. Song, H. Chen, J. Cheng,  C. Li and Y. Liu are with the Department
of Biomedical Engineering, Hefei University of Technology, Hefei
230009, China (e-mail: rcsong@hfut.edu.cn; 2018110057@mail.hfut.edu.cn; chengjuan@hfut.edu.cn; changli@hfut.edu.cn; yuliu@hfut.edu.cn ).}
\thanks{X. Chen is with  the Department of Electronic Engineering \& Information Science, University of Science and Technology of China, Hefei 230026, China  (e-mail: xunchen@ustc.edu.cn).}
}

\maketitle

\begin{abstract}
Remote photoplethysmography (rPPG) is a non-contact technique for measuring cardiac  signals from facial videos. High-quality rPPG pulse signals are urgently demanded in many fields, such as health monitoring and emotion recognition. However, most of the existing rPPG methods can only be used to get average heart rate (HR) values due to the limitation of inaccurate pulse signals.   In this paper,   a new framework based on generative adversarial network, called PulseGAN, is introduced to generate realistic rPPG pulse signals through denoising the chrominance signals. Considering that the cardiac signal is quasi-periodic and has apparent time-frequency characteristics, the error losses defined in time and spectrum domains are both employed  with the adversarial loss to enforce the model generating accurate pulse waveforms  as its  reference.  The proposed framework is tested on the public UBFC-RPPG database in both within-database and cross-database configurations. The results show that the PulseGAN framework can effectively improve the waveform quality, thereby  enhancing the  accuracy of HR, the heart rate variability (HRV) and the interbeat interval (IBI). The proposed method achieves the best performance compared to the denoising autoencoder (DAE) and CHROM, with the mean absolute error of AVNN (the average of all normal-to-normal  intervals) improving 20.85\% and 41.19\%,  and the  mean absolute error of SDNN (the standard deviation of all NN intervals) improving 20.28\% and 37.53\%, respectively, in the cross-database test.  This framework  can be easily extended to other existing deep learning-based rPPG methods, which is expected to expand the application scope of rPPG techniques.

\end{abstract}

\begin{IEEEkeywords}
Heart rate estimation, remote photoplethysmography, generative adversarial network, pulse waveform, heart rate variability
\end{IEEEkeywords}

\section{Introduction}
\IEEEPARstart{C}{ardiac}   signal is an important physiological signal to monitor the human body's health and emotional status. The common ways for obtaining cardiac signals include electrocardiogram (ECG)  and photoplethysmography (PPG).  Both of them rely on specific sensors to contact with skins of subjects, which may be uncomfortable or unsuitable for people with sensitive skins \cite{sun2016ppg}. In recent years, there is a trend  to develop non-contact heart rate measurements through the microwave Doppler or computer vision techniques.  The remote photoplethysmography (rPPG) \cite{verkruysse2008remote} is a kind of computer vision based technique to record  color changes of  facial skins caused by corresponding heartbeats using  consumer-level cameras.

After years of development, a variety of rPPG methods have been introduced according to different  assumptions and mechanisms \cite{chen2018video}. For example,   blind source separation (BSS) \cite{poh2010non} based methods are proposed under some specific statistical assumption, while the model-based rPPG methods \cite{de2013robust,wang2017algorithmic} are derived from a skin optical reflection model. However, the assumptions of conventional methods usually cannot be fully   met in realistic situations, and   the accuracy of pulse signal extraction is limited.
This causes a difficulty to calculate reliable heart rate (HR) feature information,  especially for features like heart rate variability  (HRV) that require  high-quality  waveforms.   The conventional methods usually aim to calculate the average HR values by calculating the dominate frequency of pulse signals.

However, there is a growing demand for more diverse cardiac  features   in rPPG applications,  such as stress detection, emotional classification, and health monitoring, etc.  For example, HRV is the variation of HR cycles. It is a valuable predictor of sudden cardiac death and arrhythmic events. The spectral component of HRV can also reflect the activities of the parasympathetic  and  sympathetic nervous systems. Currently, these diverse cardiac features can usually be  obtained from high-quality pulse waveforms measured by contact ECG  or PPG. Therefore, it is urgent to develop  rPPG technology which can extract  accurate pulse waveform for calculating more physiological characteristics.

On the other hand, inspired by the rapid development of deep learning (DL) techniques, DL-based rPPG algorithms \cite{chen2018deepphys,spet2018Visual,niu2018synrhythm,niu2019robust,yu2019recovering,bian2019an,slapnicar2019contact} have also been proposed in recent years. The rPPG approaches based on DL can be generally divided into two types, the end-to-end type and the feature-decoder type.  The former ones directly establish the  mapping from  video frames to the target HR  values or pulse signals, while the latter ones get the HR targets through decoding the latent  information preprocessed from video frames.  Since DL is data-driven and  neural networks have strong fitting capabilities, the results of  DL-based rPPG methods  often outperform the conventional ones, which  inspires us to extract rPPG pulse waveforms under a DL framework.

The extraction of rPPG pulse waveform can be considered as a generative problem from the perspective of  generative models.
Since firstly proposed by Ian in 2014,  generative adversarial networks (GAN) \cite{Goodfellow2014GAN} has become the mainstream generative method  due to its  state-of-the-art performance, especially  in image processing and computer vision areas. The GAN is  consisted of two neural networks, the generator $G$ and the discriminator $D$.   The two networks are trained in an adversarial way, where $G$ generates a fake target signal to confuse the discriminator,  and $D$  makes judgments on the generated signals from the real ones, thereby prompting the results of $G$  to be closer to the references.  With the  rapid development of GAN, it has also been applied to  denoise one-dimensional signals, such as speech signals \cite{pascual2017segan,Xiang2018peech}, and   ECG signals \cite{wang2019adversarial}.  These studies enlighten us to  acquire reliable rPPG waveforms  using GAN models.

In this paper, we  propose a new framework, named as PulseGAN, to  extract  rPPG pulse signal with a conditional GAN (cGAN) \cite{Mehdi2014Conditional}. The  rough pulse signal derived from CHROM method \cite{de2013robust} is taken as the input of generator $G$,  and  the PPG signal synchronously recorded by a pulse oximeter is used as a reference.   The discriminator $D$ judges the generated signal from the reference one, where the rough input  of $G$  is taken as  a conditioning. Considering the apparent characteristics of pulse signal, besides the   adversarial loss,  we also combine the waveform  error loss in the time domain and the spectrum error loss in the frequency domain to enforce a match between the generated waveform and its reference.   Through the adversarial training between $G$ and $D$, the generator  learns to construct a rPPG pulse  as close as its ground truth.  The proposed method is tested on a public UBFC-RPPG database \cite{bobbia2017unsupervised} in two scenarios, including both within- and cross-database cases. The test results  reveal that the PulseGAN  effectively improves the accuracy of the  HR, the HRV and the interbeat interval (IBI) indexes.

In summary, the main contribution of this paper is that we introduce a PulseGAN framework  to extract realistic rPPG pulse waveforms from  rough input signals derived by some conventional method. The high-quality  waveform  makes it possible to further calculate reliable cardiac features like HRV, which can  potentially  expand the application scopes of rPPG techniques.  The framework effectively combines the benefits of conventional methods  and GAN.  The generator is enforced to learn features of reference PPG signals through  error losses defined in both time  and spectrum  domains in addition to  the adversarial loss. The PulseGAN framework and  related loss functions can also be easily integrated by  some other DL-based rPPG methods to further improve their performance.

\section{Related Work}
In 2008, Verkruysse \textit{et al.} \cite{verkruysse2008remote}  first verified the validity of rPPG  for HR estimation from facial videos.  They demonstrated that the green channel signal extracted from skin pixels contained strong pulsating information.  Since then, a variety of rPPG methods have been proposed. Among them, the typical ones include those methods based on blind source separation (BSS) or the skin optical reflection model.  The BSS method assumes that the pulse signal is linearly mixed with   other noise signals, and all those signals satisfy some statistical property.  For example,  Poh \textit{et al.} \cite{poh2010non} applied independent component analysis (ICA) to separate the pulse signals from the color RGB signals. Wei \textit{et al.} \cite{wei2017non}  employed the second-order blind source separation to  extract the target signal from  six RGB channels obtained in two facial regions of interest (ROIs). On the other hand, the methods based on the optical reflection model extract pulse signal explicitly through a combination of individual color channels are combined with specific   ratios. This is considered to   eliminate  the common interference sources from the RGB channels. For example, De Haan \textit{et al.} \cite{de2013robust} proposed a chrominance method  (CHROM) to calculate the pulse signal. The CHROM method  eliminates the specular reflection component with a projection and then obtains the pulse through an "alpha tuning".   In \cite{wang2017algorithmic}, Wang \textit{et al.} used a different projection  plane orthogonal to skin color (POS) for rPPG signal extraction.  These conventional methods have achieved excellent results in calculating the average HR values of rPPG, during both laboratory and realistic scenarios. However, the quality of the waveforms remains poor  due to noise interference and model limitations, which still has large room for improvement.

\begin{figure*}[t]
\centering
\includegraphics[width=0.9\textwidth]{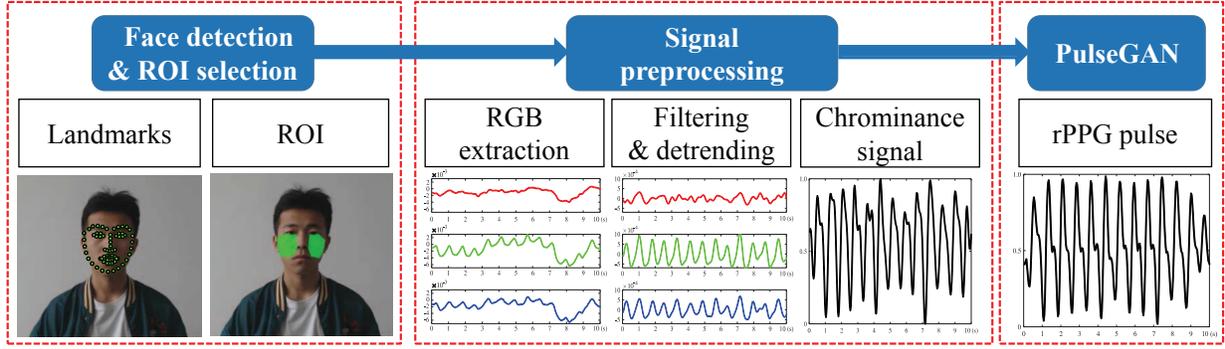}
\caption{The framework of the proposed PulseGAN method.}
\label{flowchart}
\end{figure*}

In the last few years,  DL techniques have been  increasingly used in  rPPG extraction. Here we  list some typical methods. In 2018, Chen \textit{et al.} \cite{chen2018deepphys} introduced an end-to-end system to obtain HR and respiration rate.  A convolutional neural network (CNN) combined with an attention mechanism was designed to establish the mapping between video frames and the desired physiological information.  In the same year, \v{S}petl¨ªk \textit{et al.} \cite{spet2018Visual} put forward a two-step CNN composed by a feature extractor and an HR estimator to estimate the HR from a series of facial images. Niu \textit{et al.} \cite{niu2018synrhythm} proposed a spatiotemporal representation of HR information and designed a general-to-special transfer learning strategy to estimate HR from the representation. Later, the authors also applied a channel and spatial-temporal attention mechanism to further  improve the HR estimation  from face videos \cite{niu2019robust}. Song \textit{et al.}  \cite{Song2020Heart} designed a feature-decoder framework to map a novel spatiotemporal map to the corresponding HR value through a CNN.  They also took a transfer learning to reduce the demand of training data and accelerate the convergence of model.

The goal of  above  DL-based rPPG methods is to determine   accurate HR values.  There are also some DL methods that can directly generate pulse waveforms.  For example, Bian \textit{et al.} \cite{bian2019an} proposed a new regression model that used a two-layer long short-term memory (LSTM) to filter the noisy rPPG signals. Slapni\v{c}ar \textit{et al.} \cite{slapnicar2019contact} also employed a  LSTM model to enhance the rough rPPG signals obtained by the POS algorithm.  In \cite {yu2019recovering}, Yu \textit{et al.} introduced an end-to-end way to extract pulse signal with deep spatial-temporal convolutional networks from the original face sequences. Particularly,  the authors also calculate the HRV features to evaluate the quality of pulse waveforms. Although these articles have made significant progresses in extracting waveforms, there is still room for  further improvement.

This paper aims to introduce a new framework to generate pulse waveforms with cGAN. We will verify that the proposed PulseGAN framework  improves the quality of waveforms much better than that of using the generator network with only a waveform loss.  By this means, the proposed PulseGAN framework can also be integrated into some existing methods to further improve their performance of generating  pulse waveforms.

\section{Method}
In this section, we introduce the details of the proposed PulseGAN  framework for cardiac pulse extraction.  The overall framework of PulseGAN is shown in Fig.\ref{flowchart}. First,  68-point  facial landmarks  \cite{zhang2014facial}  are detected and a region of interest (ROI) is defined according to those landmarks covering  the left and right cheeks.    Second, the  pixels within the selected ROI are averaged to get the RGB channels, and the CHROM algorithm is used to obtain a rough pulse signal that will be taken as the input of PulseGAN. Finally,  a high-quality pulse waveform is obtained through denoising the rough CHROM signal with the PulseGAN.

\subsection{Acquisition of rough rPPG pulses}
 A rough rPPG pulse signal is obtained with some conventional method before feeding into the PulseGAN. It can significantly simplify the training difficulty of PulseGAN if the rough rPPG pulse is close enough to its reference one. In this paper, the CHROM \cite{de2013robust} proposed by De Haan \textit{et al.} is taken to extract the rough rPPG pulse signal. Theoretically, other conventional methods can also be used. We choose the CHROM method because it is fast and stable against motion artifacts.

 The principle of CHROM is based on the skin optical reflection model \cite{wang2017algorithmic}. The chrominance signals $S_1$ and $S_2$ are defined based on a projection of standardized RGB signals to remove the specular reflection terms.  The rough pulse signal $X$ is then calculated through an alpha tuning technique as $X=S_{1,f}-\alpha S_{2,f}$, where $\alpha=\sigma\left(S_{1,f}\right) / \sigma\left (S_{2,f}\right)$, $\sigma$ indicates the standard deviation operation, and the  $S_{1,f}$ and $S_{2,f}$ are band-pass version of $S_1$ and $S_2$. To standardize all input signals, the obtained CHROM signal is de-trended and then normalized to  a range of $[0,1]$.

\subsection{The PulseGAN framework}
\begin{figure}[t]
\centering
\includegraphics[width=0.45\textwidth]{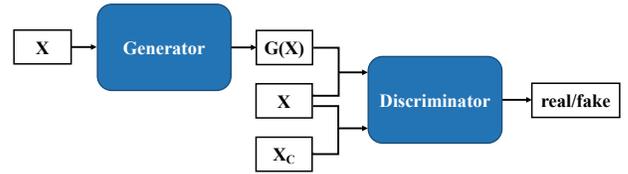}
\caption{The conditional GAN structure used in  PulseGAN.}
\label{GANframework}
\end{figure}

The overall structure of the PulseGAN  is as shown in  Fig.\ref{GANframework}. The PulseGAN is composed of a generator $G$ and a discriminator $D$. The generator $G$ is taken to map the rough CHROM signal $X$  to a target rPPG signal $G(X)$ that is close  to the reference PPG signal $X_c$. The discriminator $D$ is used to distinguish the ground truth $X_c$ from the signals $G(X)$.  To better pair the inputs and outputs, we refer to the conditional GAN \cite{Mehdi2014Conditional} approach, where the input $X$ is set as a condition in the discriminator.  Therefore, the input of the discriminator is composed of two channels  as $(G(X),X)$ or  $(X_c, X)$.
The discriminator $D$ outputs a lower score  for the input $(G(X), X)$, while it gives a higher score for the input $(X_c, X)$. The characteristics of the PPG signal are continuously learned through an adversarial learning between the generator and the discriminator, so that the output signal has a distribution as close as that of the reference PPG signal.

\begin{figure}[t]
\centering
\setcounter{subfigure}{0}
\subfigure[]{\includegraphics[width=0.45\textwidth]{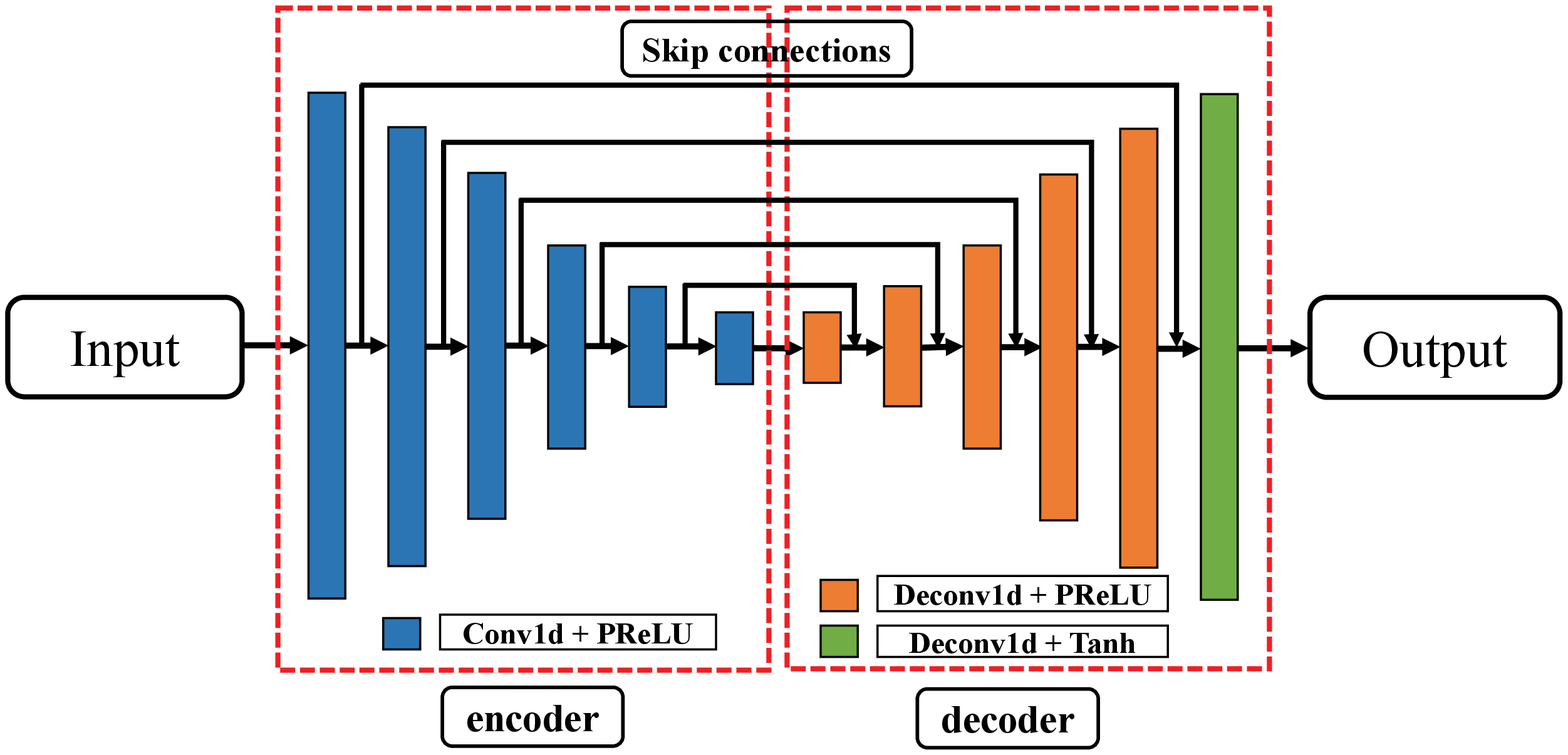}}
\hspace{8mm}
\subfigure[]{\includegraphics[width=0.36\textwidth]{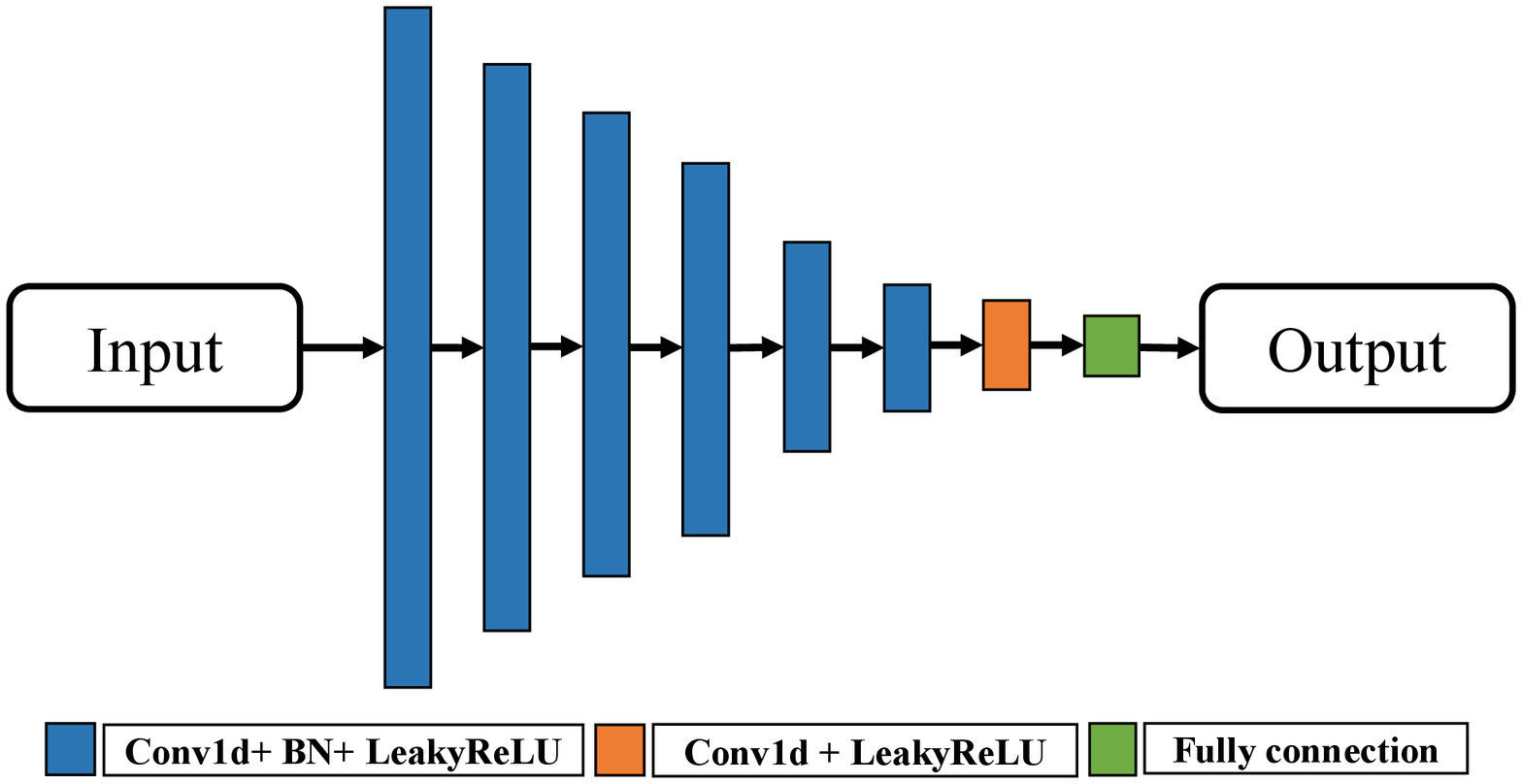}}
\caption{The network structure of PulseGAN. (a) The generator network. (b) The discriminator network. }
\label{PulseGAN}
\end{figure}

The network structures  of PulseGAN  are designed with  reference to SEGAN \cite{pascual2017segan}.  The generator, as shown in Fig.\ref{PulseGAN}(a), is similar as  a denoising  autoencoder with several skip connections. As seen, both the encoder and the decoder have six hidden layers, which are less than the ones in SEGAN.  Besides, we also remove the latent vector $z$ in SEGAN.  These modifications can reduce the risk of overfitting in generating the rPPG waveforms.   In detail, the encoder is composed of  six one-dimensional convolution layers, while the decoder has six deconvolution  layers.  The parametric rectified linear units (PReLUs)  and Tanh are taken as the nonlinear activation functions. The skip connections are taken to transfer fine-grained features from the encoder to  its counterpart in  the decoder. This is important  for the generator to   construct high-quality waveforms.


The discriminator is also a stack of several 1D convolutional layers  together with a fully connected layer in the last layer  as shown  in Fig.\ref{PulseGAN}(b). The LeakyReLU is chosen as the nonlinear activation function and  batch normalization is employed to accelerate the convergence. The input of $D$ has two channels,  where the  CHROM signal $X$ is used as a condition.  The discriminator makes judgments on the generated waveform $(G(X),X)$ and its reference one $(X_c, X)$, respectively.  The output value of $D$  represents the probability that the discriminator considers the input to be real data.


\subsection{Loss function}

The purpose of PulseGAN is to generate a waveform $G(X)$ from its input $X$.   $G(X)$ is expected to be as close as its reference signal $X_c$. This is achieved through training the PulseGAN with a lot of paired data.  Since the pulse signal has a clear time-domain and frequency-domain characteristics, we define  error losses in both domains to better guide the  generator to learn the features of the reference signal. Therefore, we define the loss function of the generator and discriminator as follows:
\begin{equation}
\label{LG}
\begin{split}
L_G = &\frac{1}{2}(D(G(X),X)-1)^2+ \lambda \parallel X_c - G(X)\parallel_1+\\
&\beta \parallel{X_c}_f - {G_f(X)}\parallel_1
\end{split}
\end{equation}
and
\begin{equation}
\begin{split}
L_D = &\frac{1}{2}(D(G(X),X))^2+ \frac{1}{2}(D(X_c,X)-1)^2.
\end{split}
\end{equation}
The first term of $L_G$ is an adversarial loss similar as the least square GAN (LSGAN) \cite{Mao2017lsgan}, the second and third ones are
  the waveform loss and the spectrum loss defined in  time domain and frequency domain, respectively. The loss function of discriminator remains the same as the LSGAN. It enforces $D$ to distinguish the generated and the reference signals. Here the ${G_f(X)}$ and ${X_c}_f$  in the spectrum loss  are  calculated by a 1024-point fast Fourier transform (FFT) on  $G(X)$ and $X_c$, respectively.   And $\parallel \cdot \parallel_1$ indicates the $L_1$ norm. The $\lambda$ and $\beta$ are the   weights of the waveform loss and the spectrum loss, respectively. The  generator is enforced to  learn the time-frequency characteristics through minimizing the error losses. Therefore, the quality of generated waveforms can be effectively improved.

\begin{figure*}[t]
\centering
\includegraphics[width=0.85\textwidth]{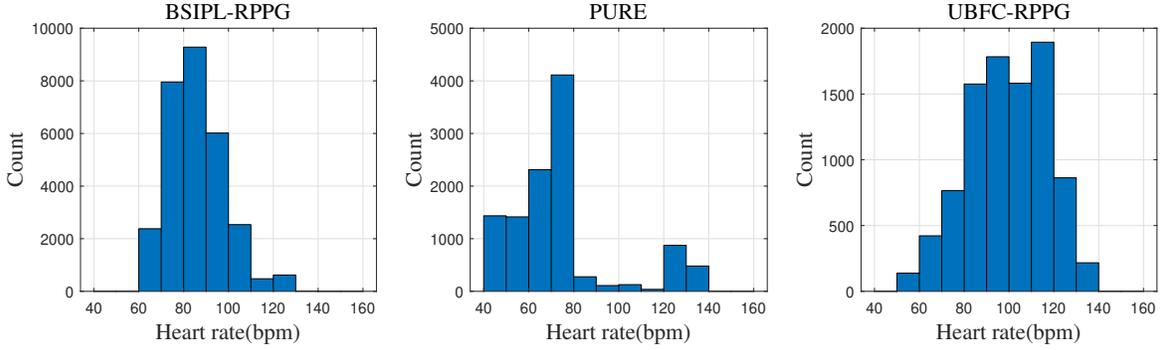}
\caption{The HR distributions of  reference PPG pulses in BSIPL-RPPG, PURE and UBFC-RPPG databases, respectively.}
\label{HRdistribution}
\end{figure*}

\section{Experiments}
In this section, we will evaluate the proposed PulseGAN   on  public  databases to illustrate its effectiveness. The PulseGAN  will be compared with several  conventional methods \cite{verkruysse2008remote, poh2010non, de2013robust, wang2017algorithmic} as well as  the  denoising autoencoder (DAE). The   conventional methods in  Table \ref{HRwithindata} have been implemented with an open source toolbox \cite{mcduff2019iphys}. The DAE here refers to use the generator $G$ of PulseGAN with only a waveform error loss.  The  quality of generated waveforms, indexed as averaged HR, the HRV, and the IBI, will be compared   under   both within-database and cross-database cases.

\subsection{Experimental setup}
To ensure the consistency of the reference waveform data from different databases, we choose to evaluate our approach with publicly available databases using the same PPG  acquisition device. This can avoid or minimize potential data errors due to differences in the PPG reference waveforms. Under this condition, three databases are selected  in our experiment including the UBFC-RPPG \cite{bobbia2017unsupervised}, the PURE \cite{Stricker2014Non-contact}, and the in-house BSIPL-RPPG databases. They all acquire the reference PPG signals  by the  Contec  CMS50E pulse oximeter. The HR distribution of each database is  shown in Fig.\ref{HRdistribution}. As can be seen, the UBFC-RPPG database has a wide range of HR distribution compared to  that of the other two. The PURE database has  a HR distribution mainly concentrating at both ends, whereas the BSIPL-RPPG database has a HR  distribution in the vicinity of around 80 bpm.

The proposed method is tested on the  UBFC-RPPG  database under two scenarios, within-database and cross-database. In order to balance the HR  distribution in training and testing sets, the PURE and BSIPL-RPPG databases are combined as training set for the cross-database case. We use a 10-second sliding window to process all videos and PPG signals for both scenarios. However, the sliding step in the within-database case is taken as   0.5 seconds, while  1 second is used for the cross-database case. A smaller sliding step can help to increase the number of training samples for the within-database case.  All reference PPG signals are resampled  to be aligned with the video frame rate.

We train  the proposed PulseGAN for 30 epochs using the Adam optimizer. The initial learning rate is set to 0.001,  and it is adaptively adjusted  through a dynamic learning rate scheduler, the 'ReduceLROnPlateau' with the  factor to 0.1  and patience to 3.  The weight parameters $\alpha$ and $\beta$ in Eq. \eqref{LG} are both taken as 10 to balance the waveform and spectrum losses.  The batch size is set to 4 in the within-database scenario, and set to 8 for the cross-database case.

\subsection{Databases}
The UBFC-RPPG  \cite{bobbia2017unsupervised} database includes 42 videos under a realistic situation.  The subjects were asked to play a time-sensitive mathematical game in order to keep the HR varied. The videos were recorded by a webcam (Logitech C920 HD Pro) with a spatial resolution of $640 \times 480$ pixels and  a frame rate of 30 fps.  Each video is about 2 minutes long, and the PPG pulse signals are collected simultaneously by the pulse oximeter (Contec Medical CMS50E) with a 60 Hz sampling rate.

The PURE \cite{Stricker2014Non-contact} database contains 60 videos from 10 subjects (8 male and 2 female). Each subject performed six different  kinds of head motions, including steady, talking, slow translation, fast translation, small rotation, and medium rotation. Each video is about 1 minute long and recorded by an ECO274CVGE camera with a resolution of $640 \times 480$ pixels and a frame rate of 30 fps. The PPG pulse signals are also collected by the Contec CMS50E  pulse oximeter while recording each video.

\begin{figure}[t]
\centering
\includegraphics[width=0.4\textwidth]{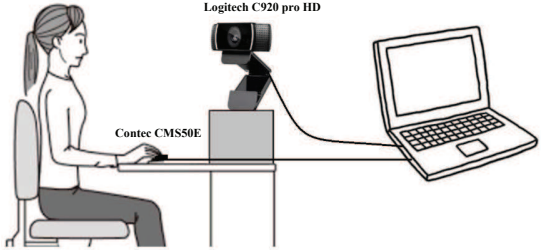}
\caption{The setup of the BSIPL-RPPG database.}
\label{BSIPL-RPPG}
\end{figure}

The BSIPL-RPPG is an in-house rPPG database  including 37 healthy student subjects (24 male and 13 female with age ranging from 18 to 25 years old). The experimental setup is illustrated in Fig.\ref{BSIPL-RPPG}. The subjects were asked to sit in front of the camera (Logitech C920 pro HD) at a distance  of 1.0 meter. A Contec CMS50E pulse oximeter was clamped on the subject's finger to acquire the PPG signal synchronously.  Both the camera and the pulse oximeter were connected to a computer to transfer the acquired data in real time.   The videos were recorded  with a resolution of $640 \times 480$ pixels under a frame rate of 30 fps. Meanwhile, the PPG signal was collected by the pulse oximeter at a 60 Hz sampling rate.  Each video and its counterpart PPG signal last about 4.5 minutes  long. The subjects were requested to sit still for the first 2 minutes, and perform some apparent head movements for the last 2.5 minutes.

\subsection{Metrics}
We define several metrics  to evaluate the quality of the generated pulse waveform. First, the  IBI sequences are calculated separately  for the generated and reference pulse signals.  A series  of cardiac features can then be defined according to the calculated IBI. For example, the average HR can be calculated from IBI  as \cite{poh2011advancements}
\begin{equation}
\mathrm{HR}=60/\overline {\mathrm{IBI}}.
\end{equation}
where $\overline {\mathrm{IBI}}$ is the average value of the IBI sequence for the  current processing window.
Similarly, we can also get HRV features  \cite{Malik1996Heart} of  AVNN and SDNN as follows,
\begin{equation}
\mathrm{AVNN} = \frac{1}{T} \sum_{i=1}^{T}\mathrm{RR}_i
\end{equation}
and
\begin{equation}
\mathrm{SDNN} = \sqrt {\frac{1}{T-1} \sum_{i=1}^{T}(\mathrm{RR}_i - \mathrm{AVNN})},
\end{equation}
where AVNN indicates the average of all normal-to-normal (NN) intervals, SDNN is the standard deviation of all NN intervals,  $\mathrm{RR}_i$ represents the $i$-th R-R interval, and $T$ is the total number of R-R intervals.

 Finally, we define the following error metrics to compare the HR, HRV (AVNN and SDNN), and IBI calculated from the PulseGAN and the reference signals.
 \begin{enumerate}
   \item \textbf{HR}:  The metrics of HR values include the mean  absolute error $\mathrm{HR}_{mae}$,  the root mean square error $\mathrm{HR}_{r m s e}$, the mean error rate percentage $\mathrm{H R}_{m e r}$, and the  Pearson¡¯s correlation coefficient $r$. The formulas of these metrics refer to \cite{Song2020Heart}.
   \item \textbf{HRV}: The mean absolute error of AVNN (or SDNN) is calculated as below:
\begin{equation}
Y_{mae} = \frac{1}{N} \sum_{n=1}^{N}\left|Y_n^{'}-Y_n\right|,
\end{equation}
where $Y_n$   indicates the  AVNN (or SDNN) for the $n$th window calculated from PulseGAN,  $Y_n^{'}$  is  the AVNN (or SDNN) from its reference PPG signal, and $N$ is the total number of time windows.
   \item   \textbf{IBI}: We also define metrics to evaluate the quality of  IBI directly. Since the length of the IBI vectors may be different, we refer to a similar way in \cite{liu2020detecting} to solve this issue. Namely, each IBI vector is expanded to the same length as the PPG signal. We  pad the $i$th RR interval of the IBI sequence with   values all equal to $\mathrm{RR}_i$. After the padding operation, we define the absolute error $\mathrm{IBI}_{ae}^{(n)}$  for the $n$th window as below
\begin{equation}
\mathrm{IBI}_{ae}^{(n)} = \mathbb{E}(\mid{\mathrm{IBI}^{(n)}_{predict}}-{\mathrm{IBI}^{(n)}_{label}}\mid),
\end{equation}
where $\mathbb{E}$ refers to the mathematical expectation, $\mathrm{IBI}^{(n)}_{predict}$ is the padded  IBI vector of  rPPG pulse, and $\mathrm{IBI}^{(n)}_{label}$ is the  padded  IBI vector of the ground truth. Finally, a mean absolute error for  IBI vectors from all samples is calculated by
\begin{equation}
\mathrm{IBI}_{mae} = \frac{1}{N} \sum_{n=1}^{N} \mathrm{IBI}^{(n)}_{ae},
\end{equation}
where  $N$ is the total number of time windows.

 \end{enumerate}

\subsection{Experimental  results}
The experimental results are introduced following a sequence of  within-database and cross-database configurations.
\subsubsection*{\textbf{Within-database}}
We first perform the within-database testing on the UBFC-RPPG database. According to the  time window and the sliding step configuration, we totally get 4234 samples, where  we take  the 3192 samples from the first 30 subjects as the training set, and the remaining 1042 samples from the last 12 subjects as  the testing set.

\begin{figure}[t]
\centering
\setcounter{subfigure}{0}
\subfigure[]{\includegraphics[width=0.35\textwidth]{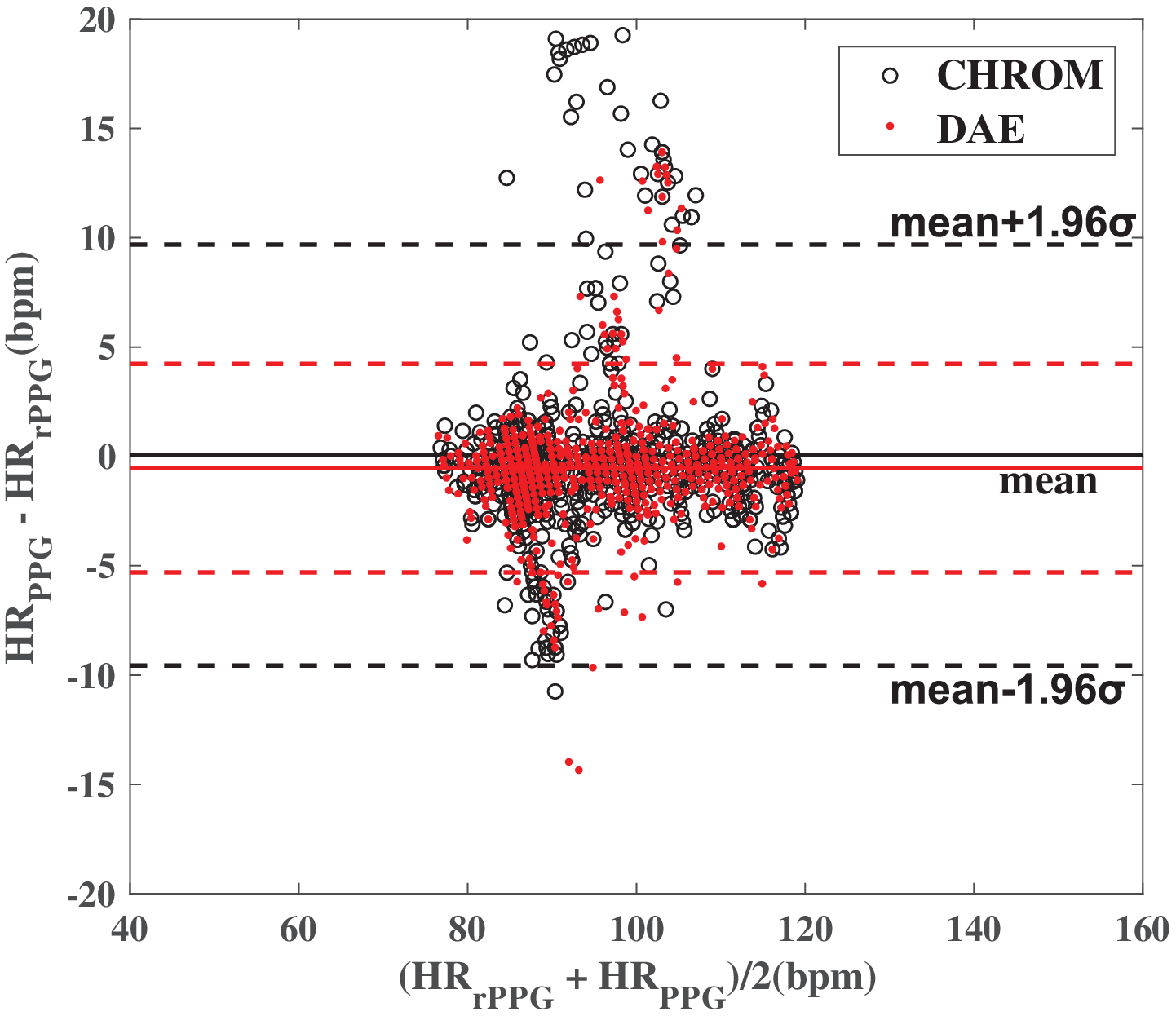}}
\hspace{8mm}
\subfigure[]{\includegraphics[width=0.35\textwidth]{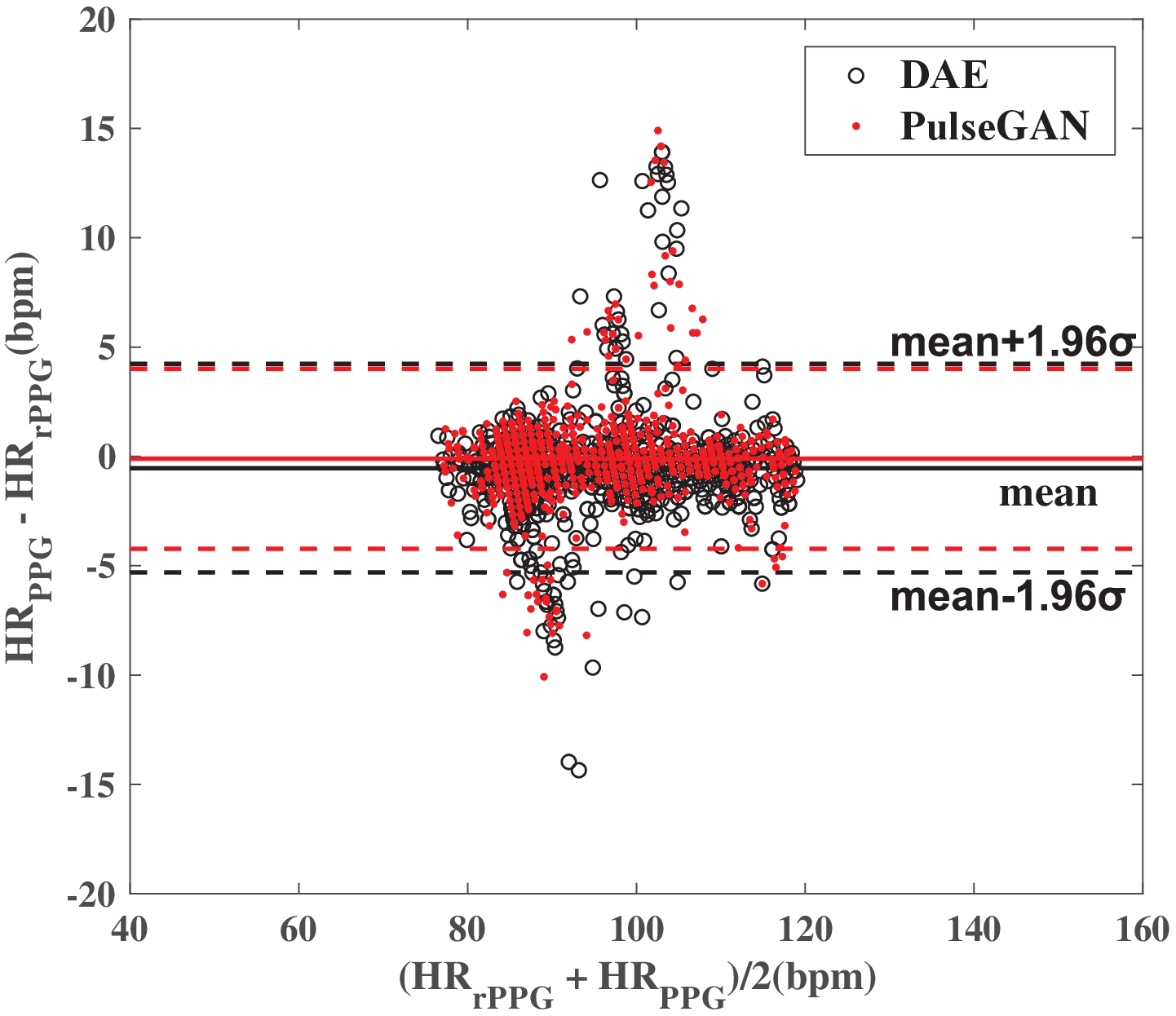}}
\caption{Bland-Altman plots between the predicted HR ($\mathrm{HR}_{\mathrm{rPPG}}$) and the reference HR ($\mathrm{HR}_{\mathrm{PPG}}$) on UBFC-RPPG database for a within-database case:  (a)  CHROM vs DAE, (b) DAE vs PulseGAN.}
\label{BA_within-database}
\end{figure}

 The estimations of average HR are summarized in Table \ref{HRwithindata}. It can be seen that the PulseGAN achieves the best performance.  The  DAE and PulseGAN can both improve the HR accuracy compared to the conventional methods. However, the PulseGAN outperforms the DAE  through the use of  the adversarial and spectrum losses.
To further compare all the results,   the Bland-Altman plots are shown in Fig.\ref{BA_within-database}. We can observe that the DAE has much better consistency with the ground truth compared to  CHROM in Fig.\ref{BA_within-database}(a).   The PulseGAN further   reduces some large errors of DAE as demonstrated  in Fig.\ref{BA_within-database}(b). The best consistency and   smallest standard deviation of the PulseGAN results indicate that the proposed method achieves the most accurate and stable  results of estimating average HR.

\newcommand{\tabincell}[2]{\begin{tabular}{@{}#1@{}}#2\end{tabular}}
\begin{table}[]
\centering
\caption{the results of average HR measurements on UBFC-RPPG database: a within-database case.}
\label{HRwithindata}
\setlength{\tabcolsep}{2.6mm}{
\begin{tabular}{|l|c|c|c|c|}
\hline  Method&
\tabincell{c}{$\mathrm{HR}_{mae}$\\(bpm)} &
\tabincell{c}{$\mathrm{HR}_{rmse}$\\(bpm)} & $\mathrm{HR}_{mer}$ & $r$ \\
\hline Verkruysse  \textit{et al}.\cite{verkruysse2008remote} & 7.50 & 14.41 & 7.82\% & 0.62 \\
\hline Poh \textit{et al}.\cite{poh2010non} & 5.17 & 11.76 & 5.30\% & 0.65 \\
\hline Wang \textit{et al}.\cite{wang2017algorithmic} & 4.05 & 8.75 & 4.21\% & 0.78 \\
\hline Haan \textit{et al}.\cite{de2013robust} & 2.37 & 4.91 & 2.46\% & 0.89 \\
\hline DAE & 1.48 & 2.49 & 1.55\% & 0.97 \\	
\hline PulseGAN & 1.19 & 2.10 & 1.24\% & 0.98 \\
\hline	
\end{tabular}}
\end{table}

\begin{figure*}[]
\centering
\setcounter{subfigure}{0}
\subfigure[]{\includegraphics[width=0.35\textwidth]{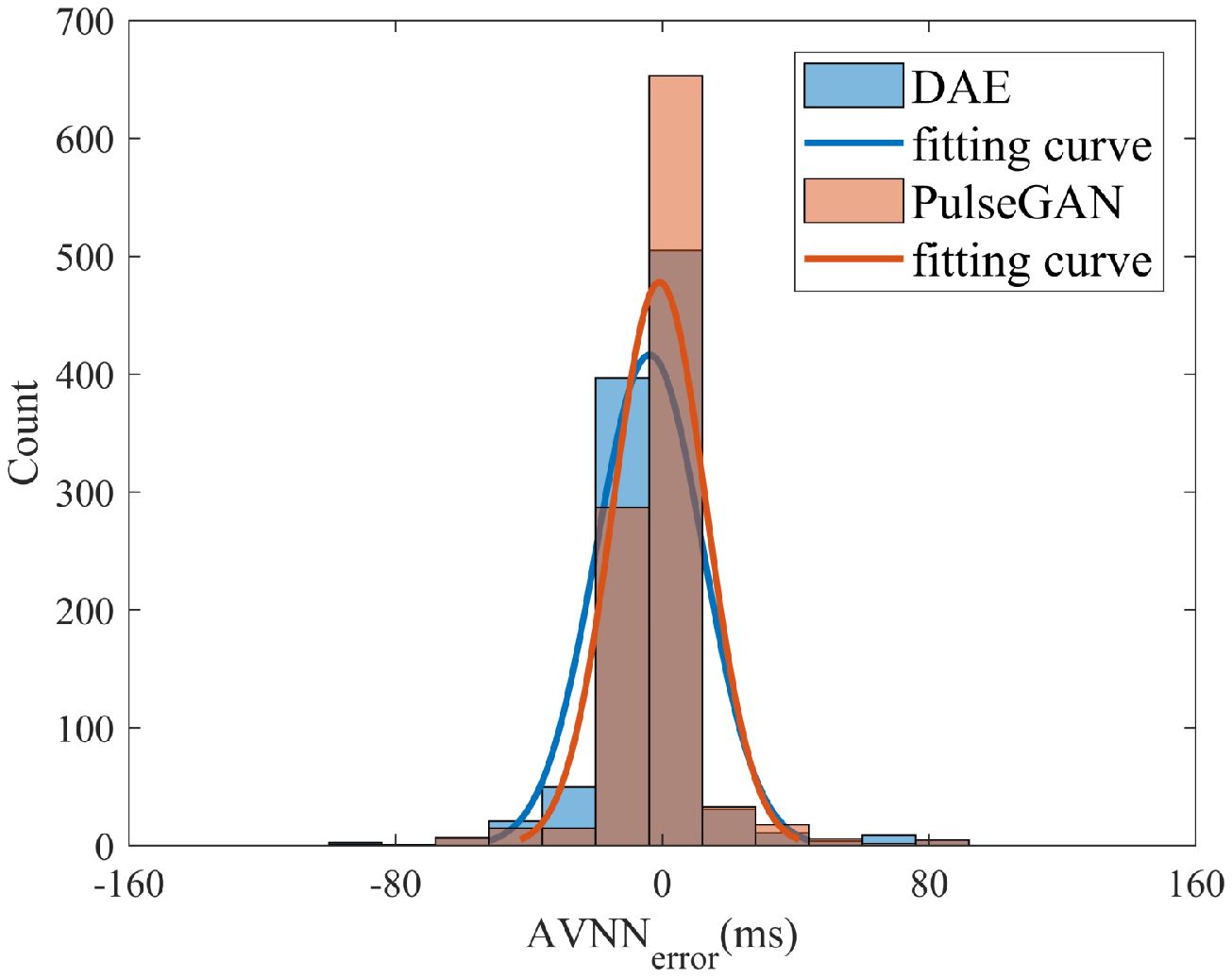}}
\hspace{8mm}
\subfigure[]{\includegraphics[width=0.35\textwidth]{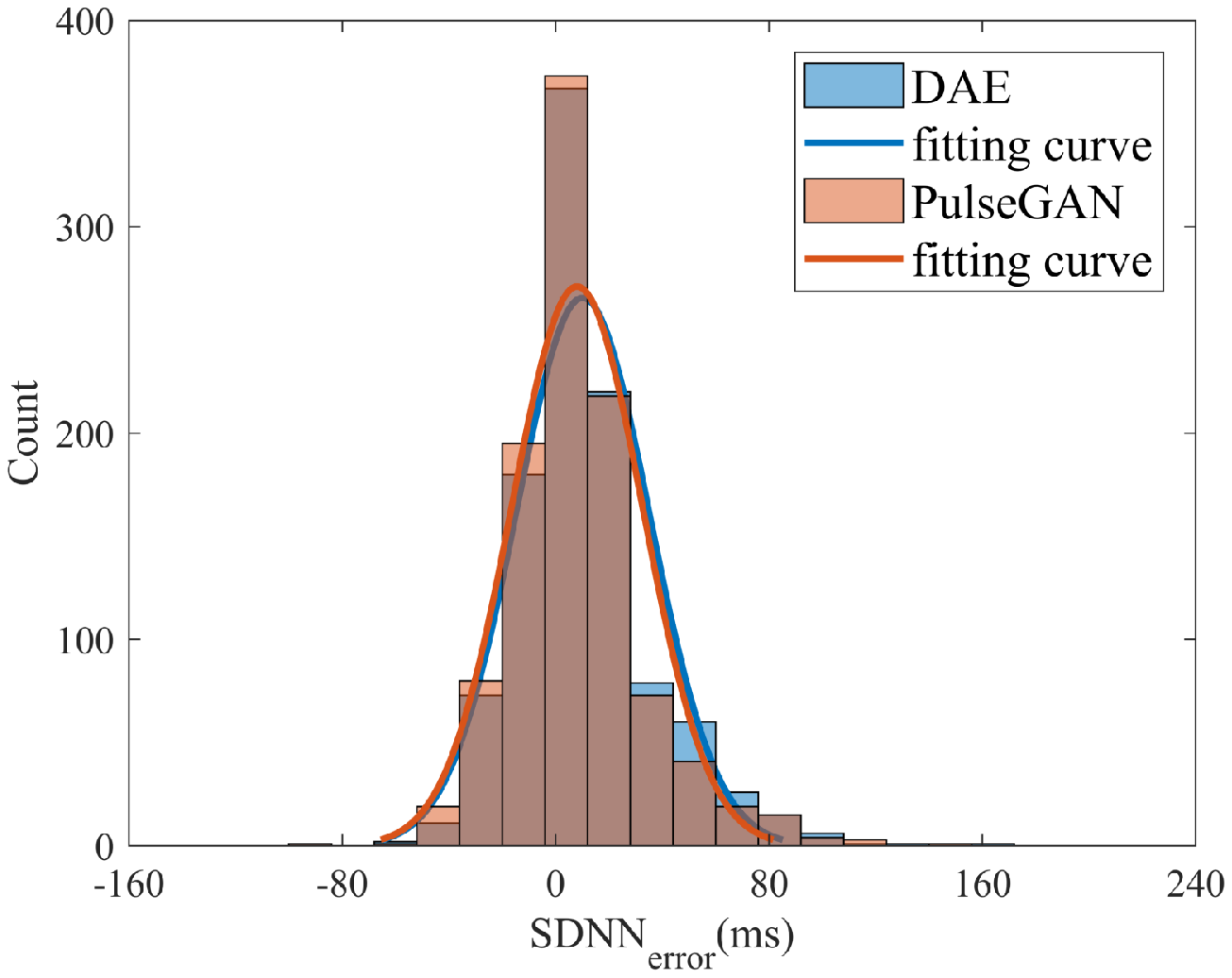}}
\hspace{8mm}
\subfigure[]{\includegraphics[width=0.35\textwidth]{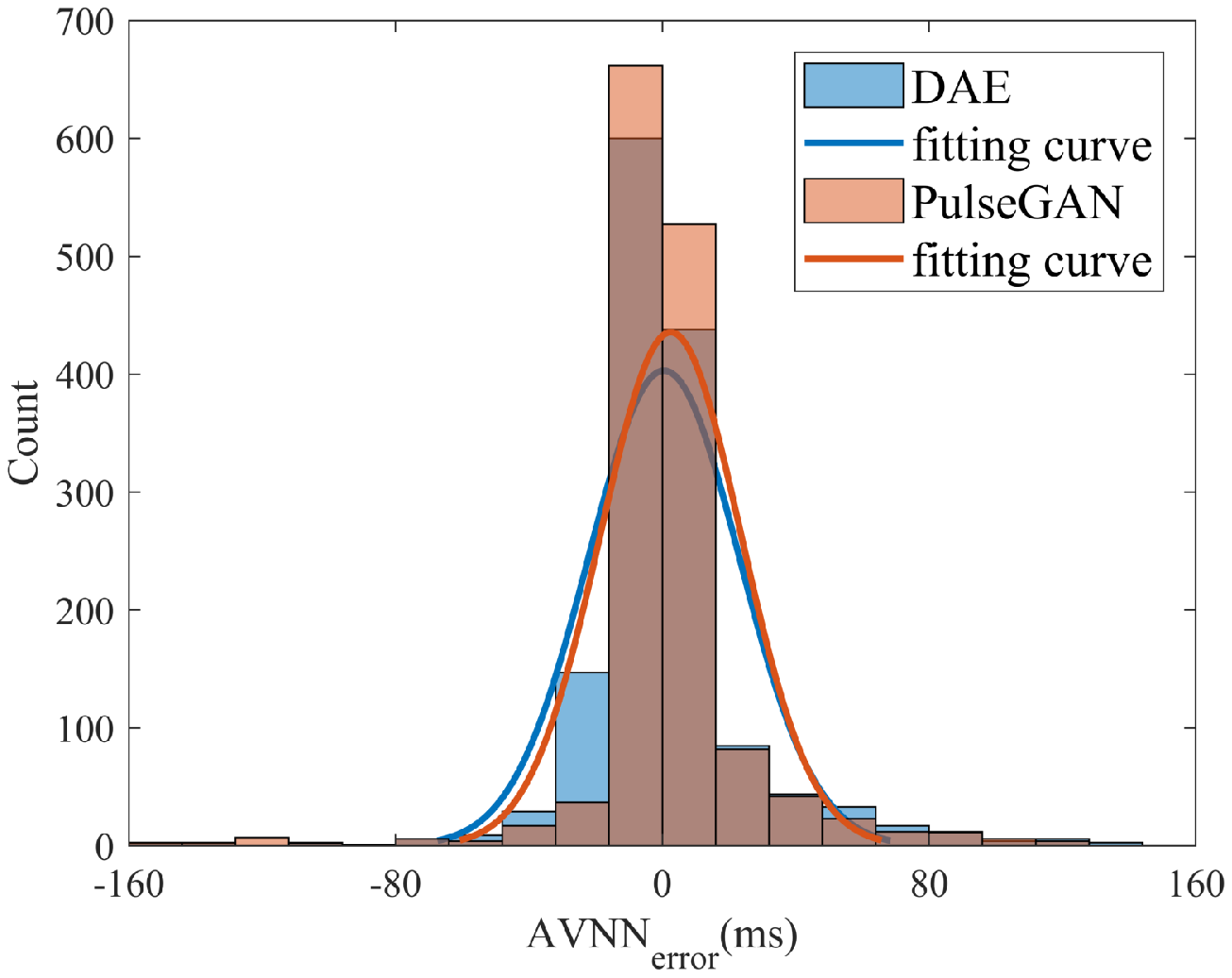}}
\hspace{8mm}
\subfigure[]{\includegraphics[width=0.35\textwidth]{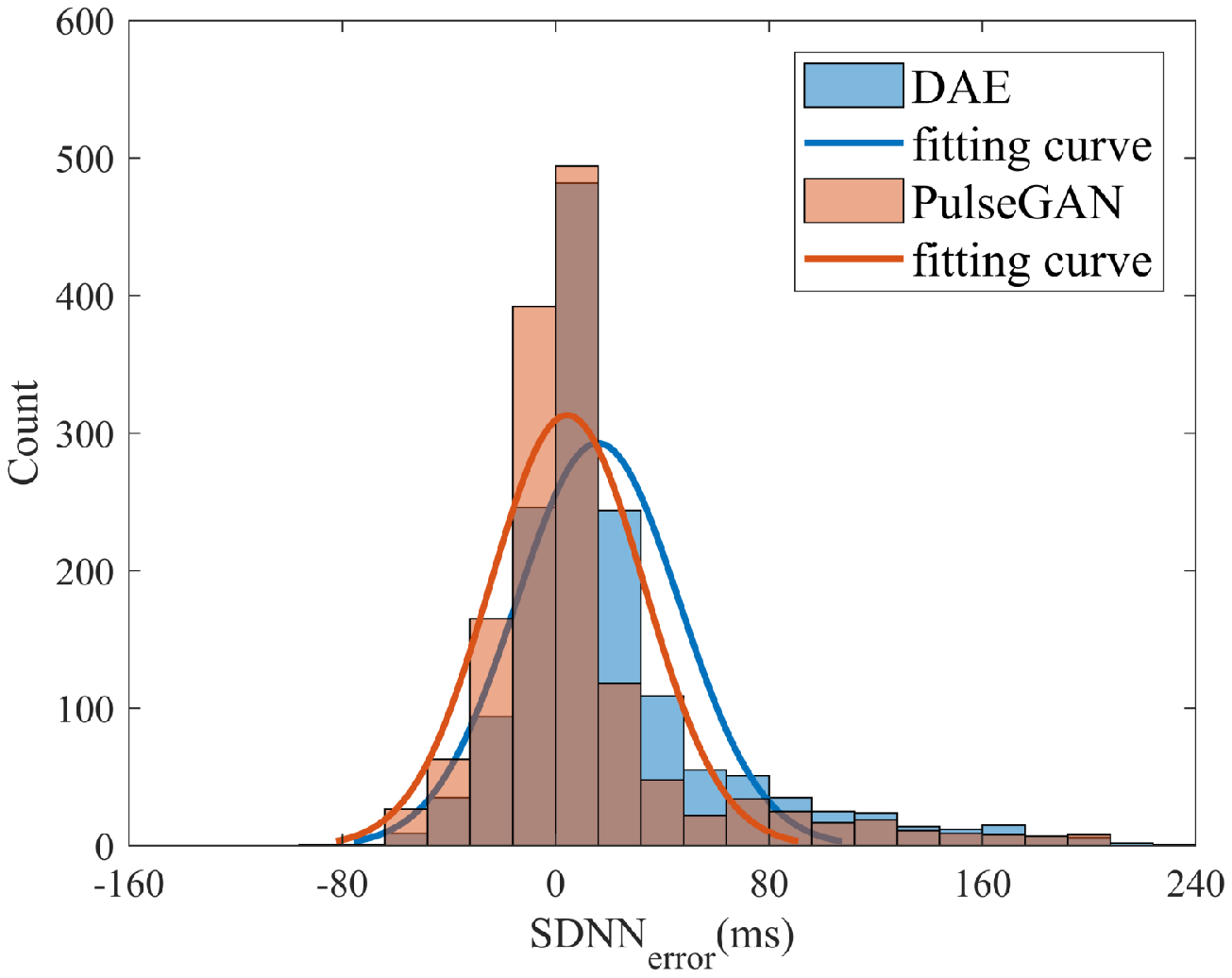}}
\caption{The statistical histograms of errors of AVNN (left) and SDNN (right). The   (a) and (b)  are for a within-database case. The  (c)  and (d) are for a  cross-database case.}
\label{HRVerror}
\end{figure*}

\begin{figure*}[t]
\centering
\setcounter{subfigure}{0}
\subfigure[]{\includegraphics[width=0.35\textwidth]{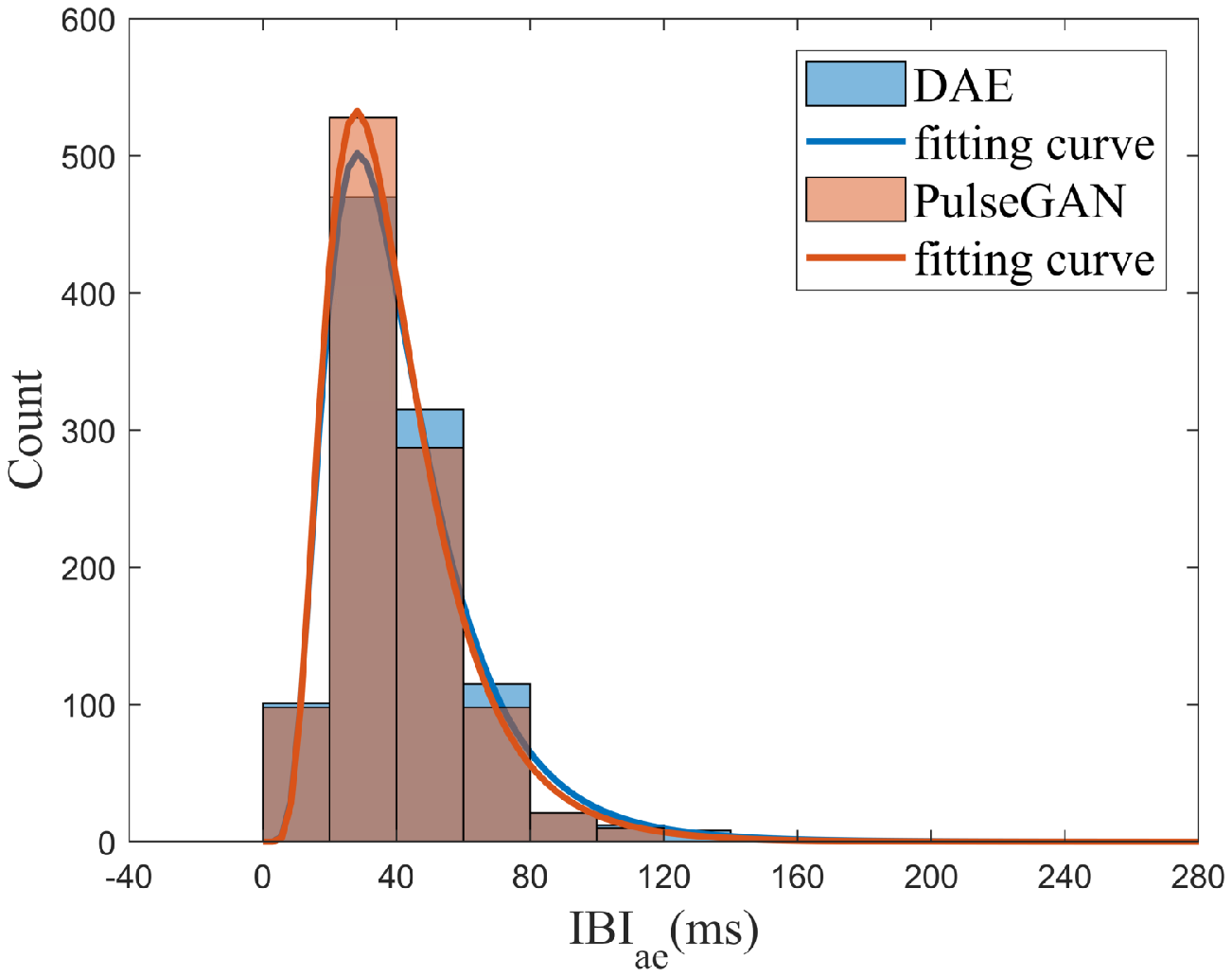}}
\hspace{8mm}
\subfigure[]{\includegraphics[width=0.35\textwidth]{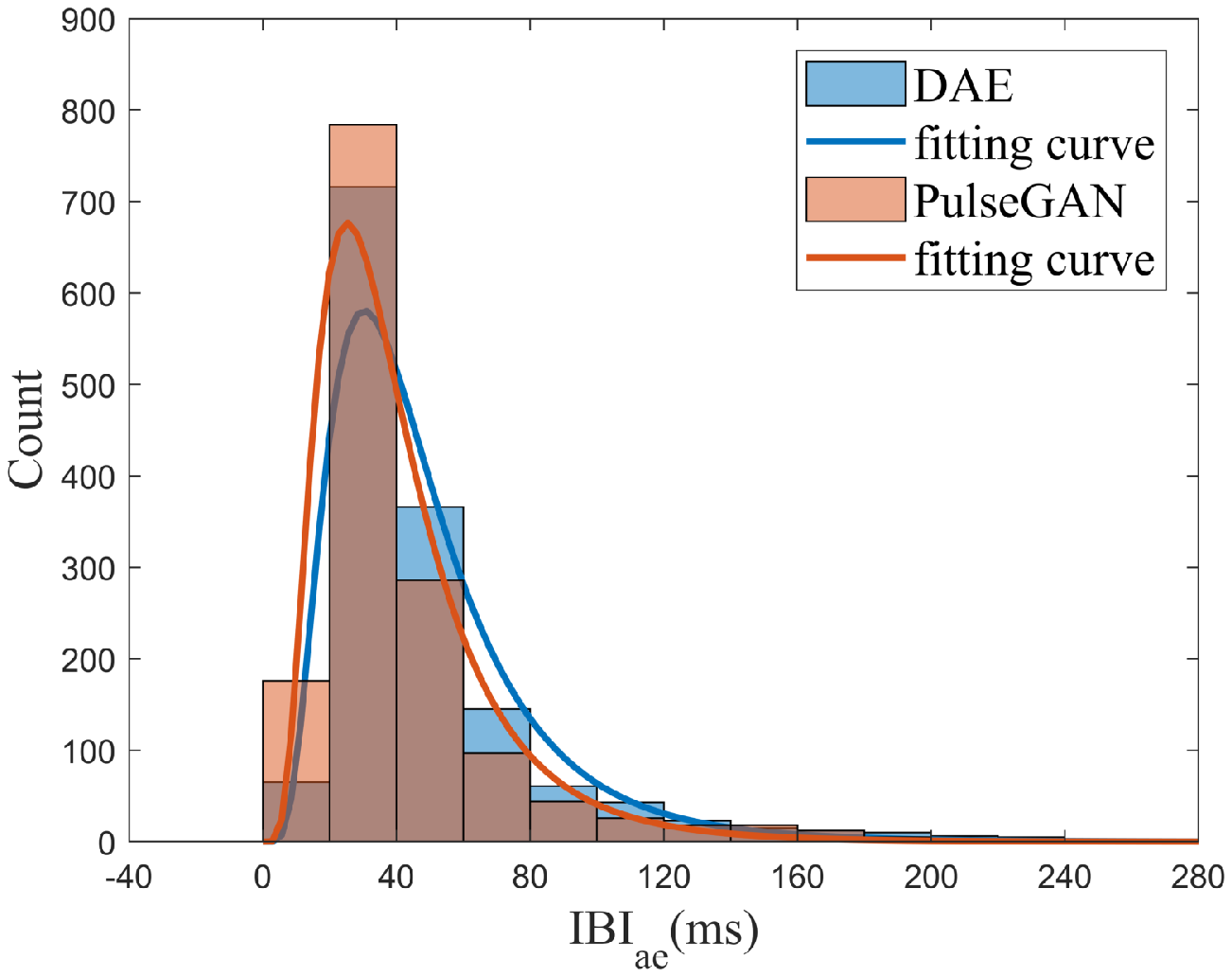}}
\caption{The statistical histograms of  absolute errors ($\mathrm{IBI}_{ae}$)  of IBI sequences: (a) within-database, (b) cross-database.}
\label{IBImae}
\end{figure*}

\begin{figure}[t]
\centering
\setcounter{subfigure}{0}
\subfigure[]{\includegraphics[width=0.45\textwidth]{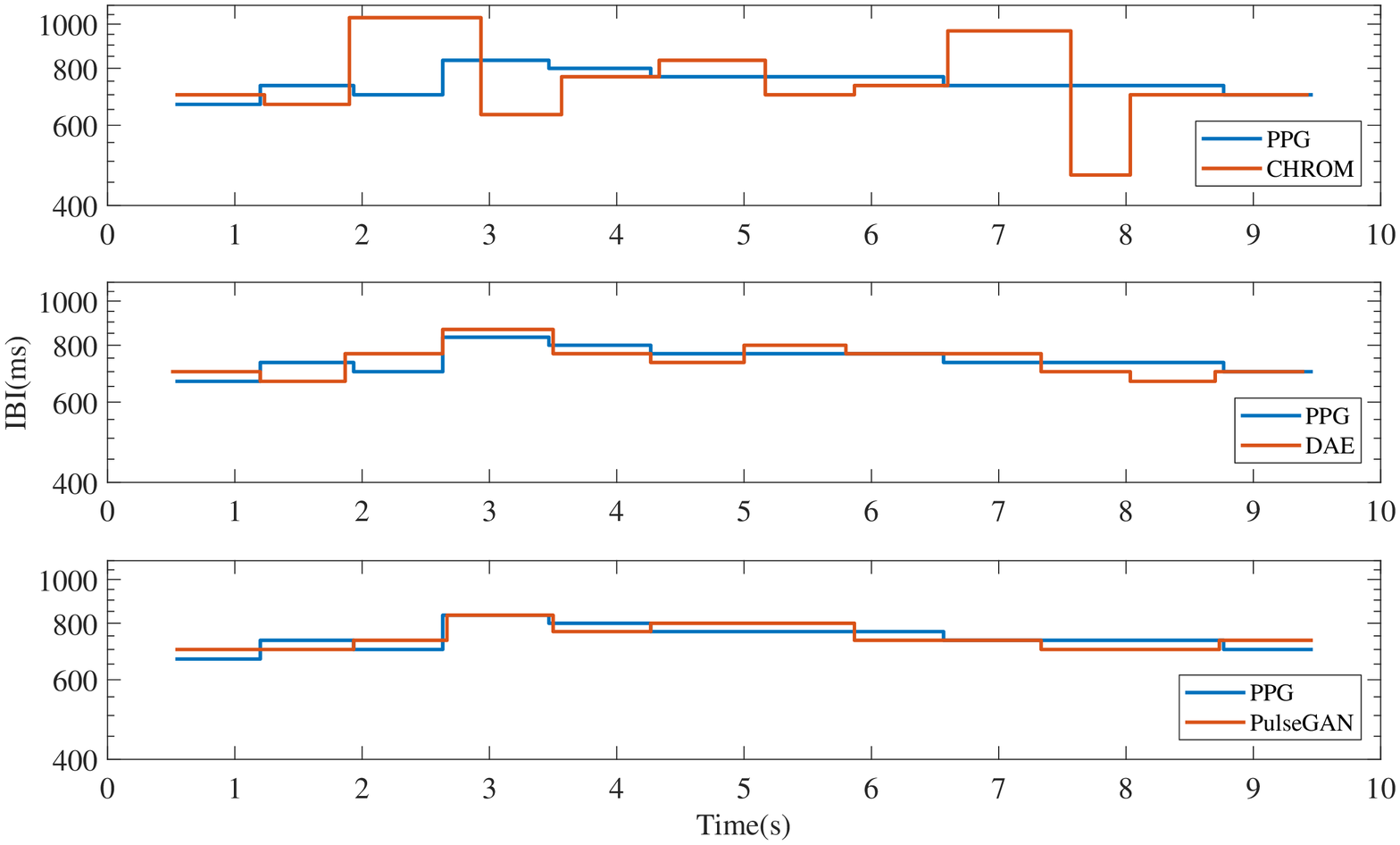}}
\hspace{8mm}
\subfigure[]{\includegraphics[width=0.45\textwidth]{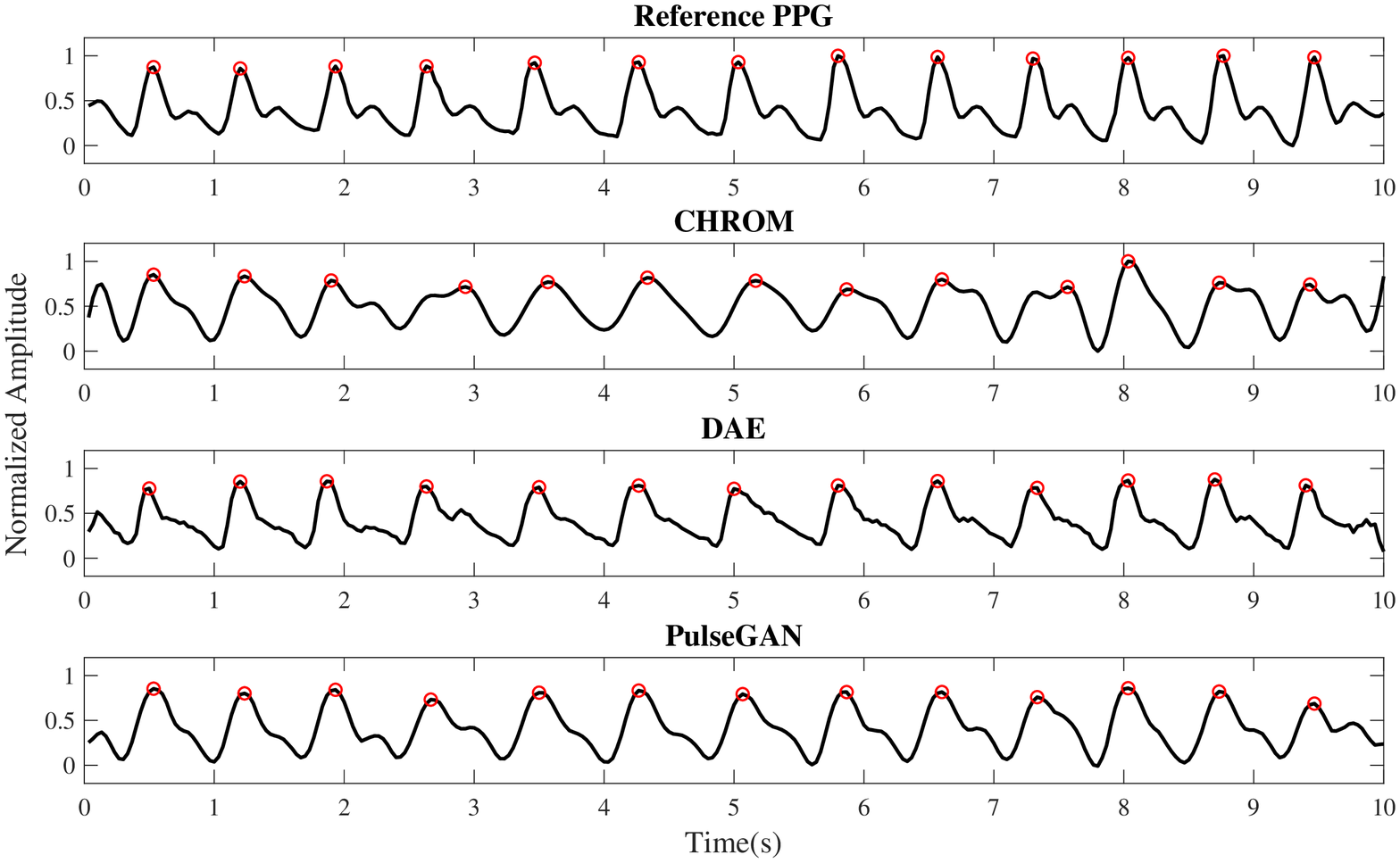}}
\caption{A comparison example of IBI sequences in (a) and rPPG pulse signals  in (b) on UBFC-RPPG database: a  within-database case.}
\label{IBI_pulse_within-database}
\end{figure}

Moreover, we  compare the performance of  CHROM, DAE and PulseGAN methods on  two HRV indexes, the AVNN and SDNN.  The results in Table \ref{HRV} show that the PulseGAN also makes a clear improvement  of HRV features in the within-database case. Compared to DAE, the PulseGAN improves both the $\mathrm{AVNN}_{mae}$ and  $\mathrm{SDNN}_{mae}$, which implies that the  waveforms generated by PulseGAN have  better qualities than those from DAE.  We also present the statistical  histograms of AVNN and SDNN errors in Fig.\ref{HRVerror}(a) and Fig.\ref{HRVerror}(b).   It can be seen that the errors of AVNN (also SDNN) by PulseGAN are more concentrated around zero than those of DAE. Similarly, the mean absolute errors of  IBI vectors (i.e., $\mathrm{IBI}_{mae}$)  are listed in Table \ref{IBI}. The error  distribution of the $\mathrm{IBI}_{ae}$  for all samples is shown in Fig.\ref{IBImae}(a). The PulseGAN  slightly improves the $\mathrm{IBI}_{mae}$ compared to DAE under the within-database configuration.   In order to observe the improvement of the waveform quality more intuitively, we demonstrate a sample of the pulse signal and corresponding IBI sequence, as shown in Fig.\ref{IBI_pulse_within-database}. It can be seen  that the waveform and IBI sequences of the example pulse signal are both significantly improved by DAE and PulseGAN compared to CHROM. The $\mathrm{IBI}_{ae}$ errors of the example in Fig. \ref{IBI_pulse_within-database}(a)  are 112.44,     42.50, and    24.67 ms for CHROM, DAE, and PulseGAN, respectively.

\begin{table}[]
\begin{center}
\centering
\caption{The mean absolute errors of AVNN and SDNN on  UBFC-RPPG database.}
\label{HRV}
\centering
\setlength{\tabcolsep}{1.2mm}{
\begin{tabular}{|l|c|c|c|c|}
\hline
\multicolumn{1}{|c|}{}&\multicolumn{4}{c|}{$\mathrm{HRV}_{mae}$(ms)}\\
\hline
\multirow{2}{*}{Method}&
\multicolumn{2}{c|}{within-database}&\multicolumn{2}{c|}{cross-database}\cr\cline{2-5}
& $\mathrm{AVNN}_{mae}$ & $\mathrm{SDNN}_{mae}$ & $\mathrm{AVNN}_{mae}$ & $\mathrm{SDNN}_{mae}$ \cr
\hline Haan \textit{et al}.\cite{de2013robust}  &16.54 &40.90 &25.30 &38.96\\
\hline DAE &9.52 &19.25 &18.80 &30.53\\
\hline PulseGAN &7.52 &18.36 &14.88 &24.34\\
\hline  	
\end{tabular}}
\end{center}
\end{table}

\begin{table}[]
\begin{center}
\centering
\caption{The mean absolute errors of IBI sequences  on UBFC-RPPG database.}
\label{IBI}
\centering
\setlength{\tabcolsep}{5.5mm}{
\begin{tabular}{|l|c|c|c|c|}
\hline
\multicolumn{1}{|c|}{}&\multicolumn{2}{c|}{$\mathrm{IBI}_{mae}$(ms)}\\
\hline Method &within-database &cross-database\\
\hline Haan \textit{et al}.\cite{de2013robust}  &63.20 &60.16\\
\hline DAE &41.27 &49.65\\
\hline PulseGAN &39.60 &42.27\\
\hline  	
\end{tabular}}
\end{center}
\end{table}

\subsubsection*{\textbf{Cross-database}}
In the case of cross-database, we  take the PURE and  BSIPL-RPPG databases as the training set. This combination can effectively balance the number of samples in different HR ranges to achieve a more consistent HR distribution with the testing set. According to the configuration of  cross-database scenario, there are total 13484 training samples  obtained  from the PURE (3727 samples) and BSIPL-RPPG (9757 samples) databases. Similarly, we get 1470 samples from the UBFC-RPPG database as the testing set.

\begin{figure}[t]
\centering
\setcounter{subfigure}{0}
\subfigure[]{\includegraphics[width=0.35\textwidth]{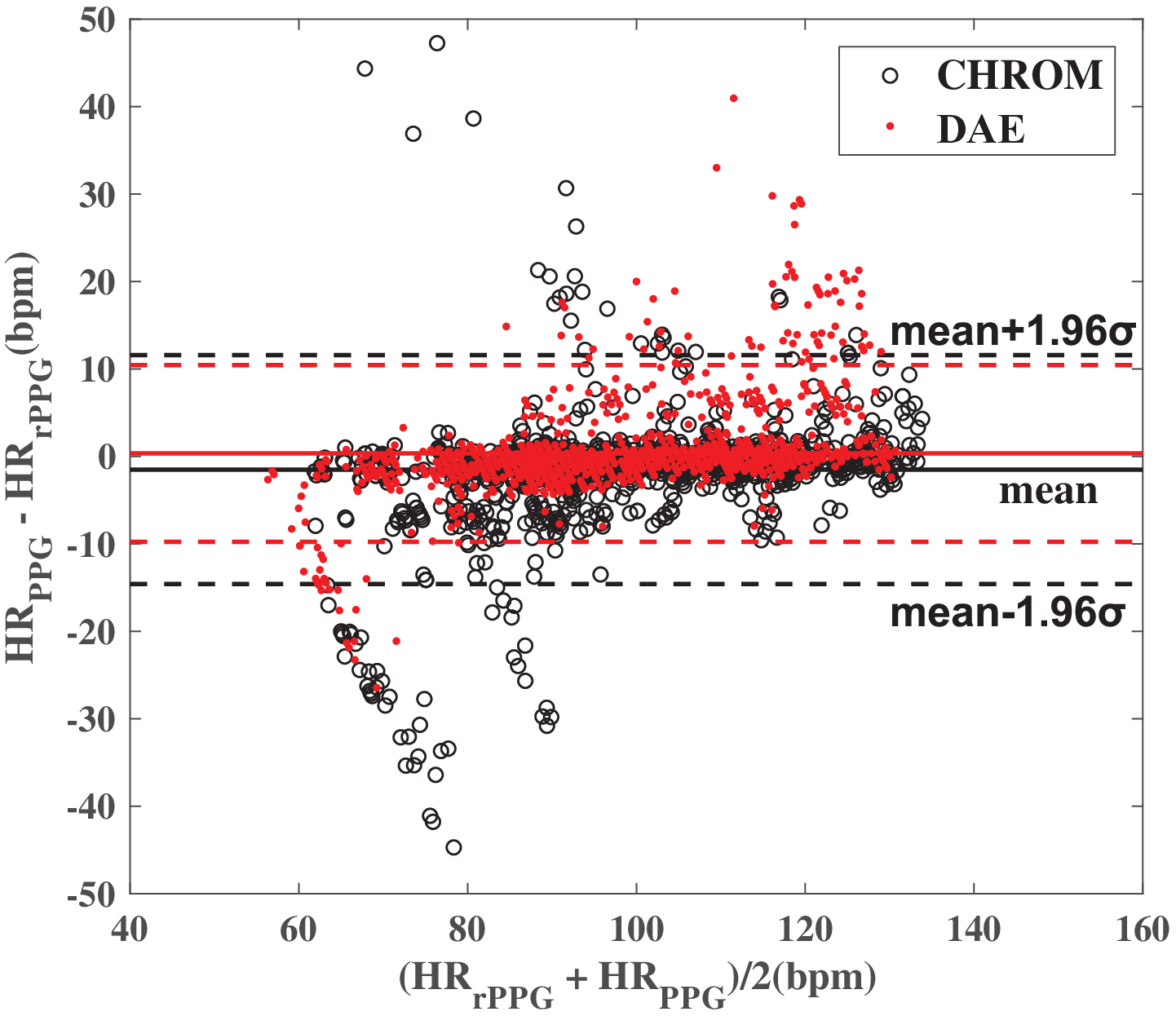}}
\hspace{8mm}
\subfigure[]{\includegraphics[width=0.35\textwidth]{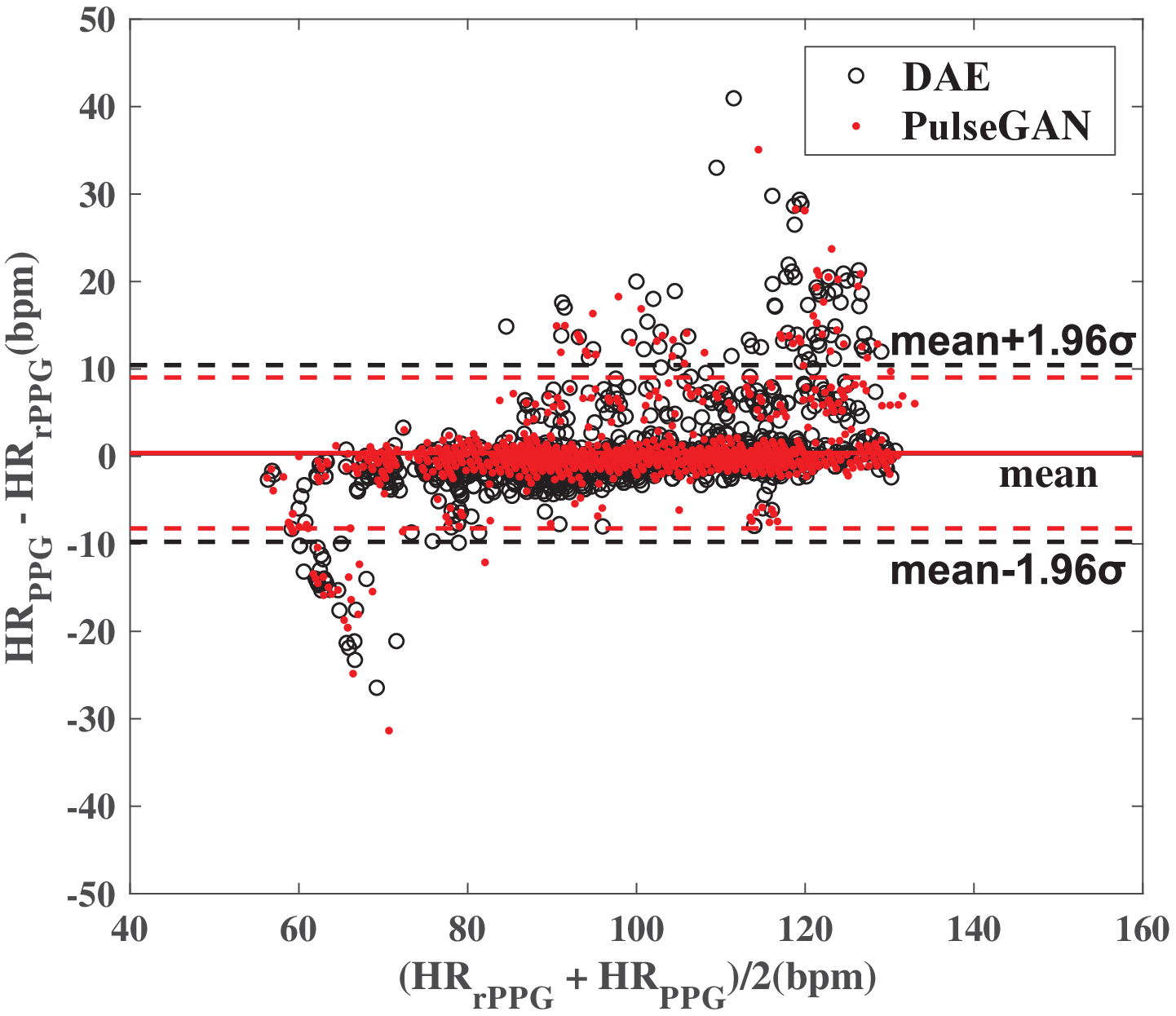}}
\caption{Bland-Altman plots between the predicted HR ($\mathrm{HR}_{\mathrm{rPPG}}$) and the reference HR ($\mathrm{HR}_{\mathrm{PPG}}$) on UBFC-RPPG database for a cross-database case:  (a)  CHROM vs DAE, (b) DAE vs PulseGAN.}
\label{BA_cross-database}
\end{figure}

 The average HR measurements are summarized in   Table \ref{HRcrossdata}.  The proposed PulseGAN  still achieves the best results compare to the other ones. Similarly, the Bland-Altman plots are illustrated in Fig.\ref{BA_cross-database}  to show the consistency of  the predicted  HR values with the reference ones.  We can see that the DAE and PulseGAN both outperform the CHROM method. The PulseGAN   achieves a even better performance compared to DAE due to its advantages in waveform reconstruction.

\begin{table}[]
\begin{center}
\caption{the results of average HR measurements on UBFC-RPPG dataset: a cross-database case.}
\centering
\label{HRcrossdata}
\setlength{\tabcolsep}{2.6mm}{
\begin{tabular}{|l|c|c|c|c|}
\hline  Method &
\tabincell{c}{$\mathrm{HR}_{mae}$\\(bpm)} &
\tabincell{c}{$\mathrm{HR}_{rmse}$\\(bpm)} & $\mathrm{HR}_{mer}$ & $r$ \\
\hline Verkruysse  \textit{et al}.\cite{verkruysse2008remote} & 8.29 & 15.82 & 7.81\% & 0.68 \\
\hline Poh \textit{et al}.\cite{poh2010non} & 4.39 & 11.60 & 4.30\% & 0.82 \\
\hline Wang \textit{et al}.\cite{wang2017algorithmic} & 3.52 & 8.38 & 3.36\% & 0.90 \\
\hline Haan \textit{et al}.\cite{de2013robust} & 3.10 & 6.84 & 3.83\% & 0.93 \\
\hline DAE & 2.70 & 5.17 & 2.85\% & 0.96 \\	
\hline PulseGAN & 2.09 & 4.42 & 2.23\% & 0.97 \\
\hline	
\end{tabular}}
\end{center}
\end{table}

\begin{figure}[t]
\centering
\setcounter{subfigure}{0}
\subfigure[]{\includegraphics[width=0.45\textwidth]{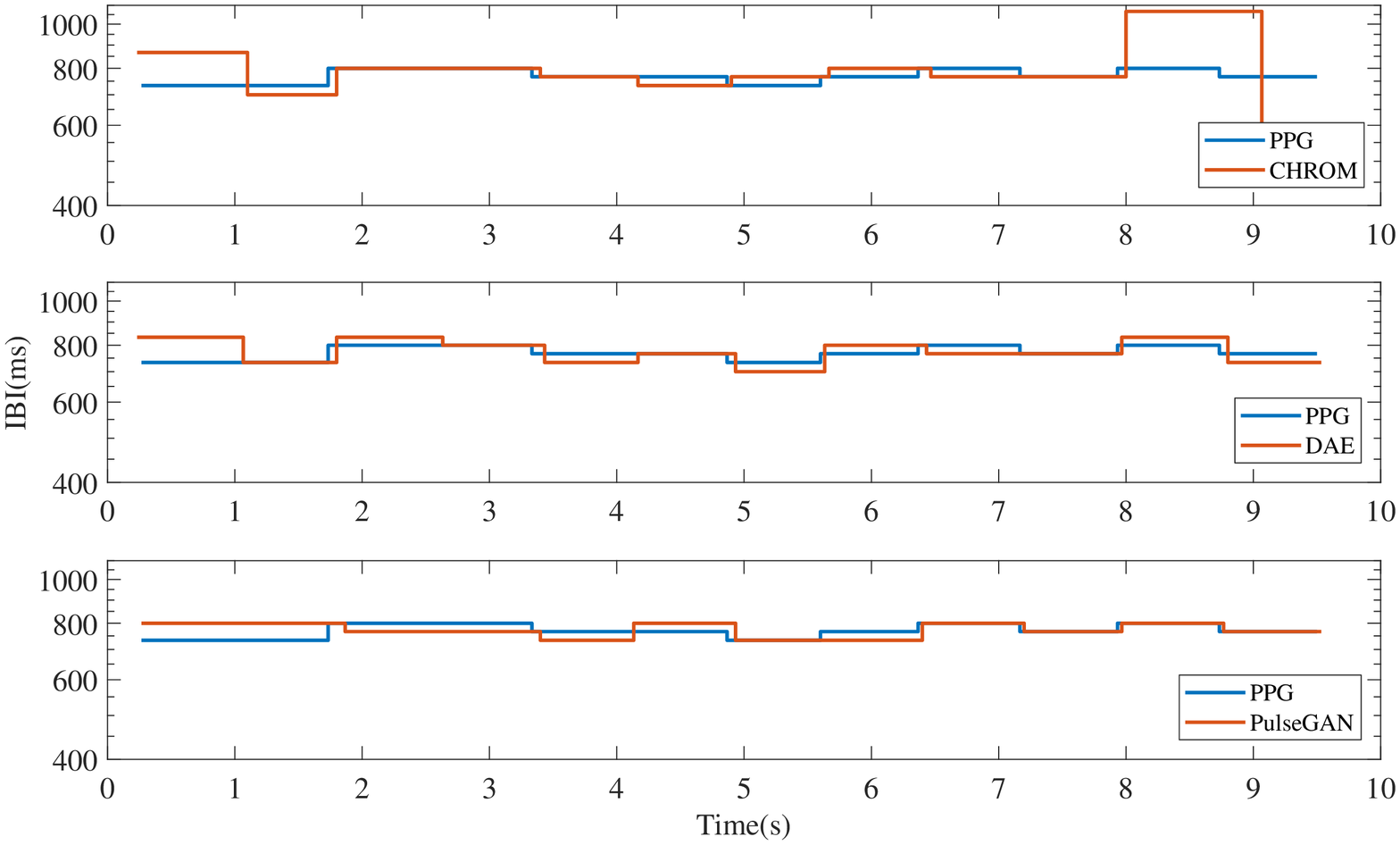}}
\hspace{8mm}
\subfigure[]{\includegraphics[width=0.45\textwidth]{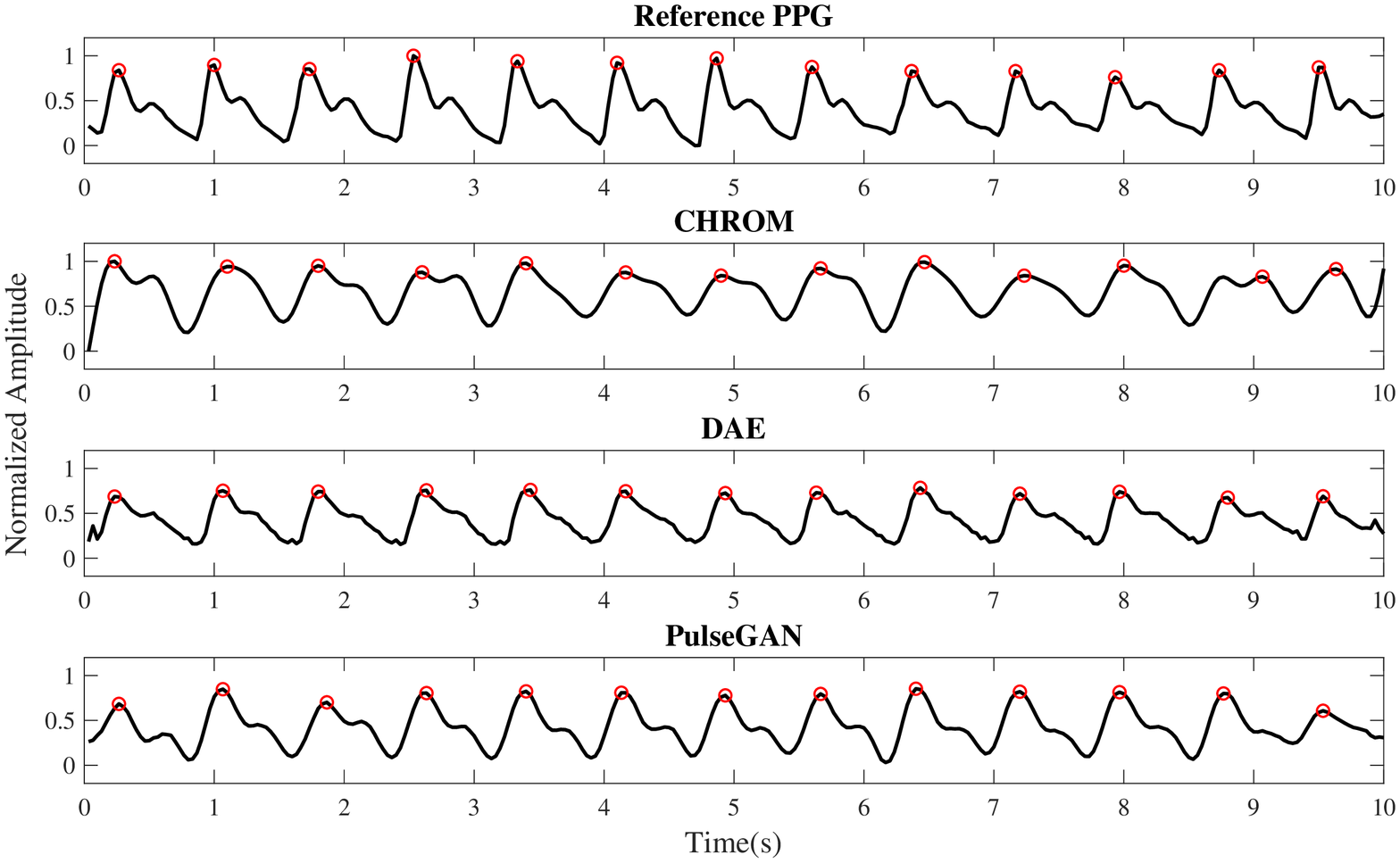}}
\caption{A comparison example of IBI sequences in (a) and rPPG pulse signals  in (b) on UBFC-RPPG database: a  cross-database case. }
\label{IBI_pulse_cross-database}
\end{figure}

The mean absolute errors of  HRV  features,  AVNN and SDNN,  are listed in Table \ref{HRV} for  the cross-database case.
 The PulseGAN achieves the best performance, with the $\mathrm{AVNN}_{mae}$  improves 20.85\% (41.19\%),  and the  $\mathrm{SDNN}_{mae}$ improves 20.28\% (37.53\%),  compared to  DAE (CHROM), in the cross-database test.
In addition, the error histogram and its distribution fitting curve are illustrated in Fig.\ref{HRVerror}(c) and (d), respectively. The results show that the PulseGAN not only improves the accuracy of heart rate but also improves the quality of HRV features in the cross-database scenario. The  mean absolute errors of IBI vectors ($\mathrm{IBI}_{mae}$) are listed   in TABLE \ref{IBI} for the cross-database case. The
 improvement of PulseGAN compared to DAE on $\mathrm{IBI}_{mae}$  is more clear  than that in the within-database case.   The error  distribution of the $\mathrm{IBI}_{ae}$  for all samples is shown in  Fig.\ref{IBImae}(b). We  observe that the PulseGAN has a remarkable improvement than the DAE.   Finally,  we take an example to demonstrate the intuitive enhancement on waveforms and the IBI vectors. As can be seen in  Fig.\ref{IBI_pulse_cross-database},   the IBI vector and the related pulse waveform obtained by PulseGAN are more close to their  ground truths compared to DAE and CHROM.  The $\mathrm{IBI}_{ae}$ errors of the example in Fig. \ref{IBI_pulse_cross-database}(a)  are 65.01,     27.44, and    23.11 ms for CHROM, DAE, and PulseGAN, respectively. The experimental results of PulseGAN for the cross-database case indicates the strong generalization capability of  the proposed model.

The above results indicate that the PulseGAN can effectively generate rPPG pulse waveforms with high qualities. It consistently outperforms the DAE method  for both   within-database and cross-database cases.  Although   the PulseGAN has not been compared with other existing deep learning-based rPPG methods, we want to emphasize that the framework of PulseGAN, including   the usage of both time-domain  and spectrum-domain losses under a GAN architecture, can be easily integrated with   these existing methods to further improve their results.

\section{Conclusion}
The cardiac pulse signal is   very important   to evaluate the healthy and emotional status of human bodies. In this paper, we have  proposed  a PulseGAN to extract high-quality  pulse waveforms through remote photoplethysmogrpahy. The PulseGAN is designed based on a framework of generative adversarial network with  error losses defined in both   time and spectrum domains.  It takes the rough CHROM signal as the input, and outputs a rPPG pulse through the deep generative model. This architecture is also easy to integrate with existing deep learning based rPPG methods and further improve their performance.  The experimental results on a public UBFC-RPPG database demonstrate that the PulseGAN consistently outperforms the DAE and other conventional methods for both  within-database and cross-database cases. The generated high-quality  waveforms from PulseGAN  have made it possible to calculate more cardiac features like AVNN and SDNN  through rPPG.   Although  the   HRV characteristics calculated in this paper are relatively preliminary,   these attempts are meaningful to expand the scope of application of rPPG.

\bibliographystyle{IEEEtran}
\bibliography{GAN-rPPG}

\end{document}